\documentclass[12pt]{article}
\usepackage{graphicx}
\usepackage{setspace}
\newcommand\beq{\begin{equation}}
\newcommand\eeq{\end{equation}}
\newcommand\bear{\begin{eqnarray}}
\newcommand\eear{\end{eqnarray}}

\begin{document}
\begin{center}

{\Large \bf Molecular Electronics: Effect of external electric field}
\end{center}

\vspace*{0.3cm}

\centerline{S. Lakshmi, Sudipta Dutta and Swapan K. Pati }

\vspace*{0.3cm}
 
\begin{center}
Theoretical Sciences Unit \\
Jawharlal Nehru Center for Advanced Scientific Research  \\
Jakkur Campus, Bangalore 560 064, India.
\end{center}
 
\vspace*{0.6cm}

\begin{abstract}

The effect of electric field, applied on systems in the nanoscale regime 
has attracted a lot of research in recent times. We highlight some
of the recent results in the field of single molecule electronics and
then move on to focus on some of our own results in this area.
We have first shown how important it is to obtain the spatial 
profile of the external bias potential across the system, and
how this would change in the presence of electron-electron interactions.
We have also studied different 
kinds of insulators in the presence of the spatially varying external bias,
and have explicitly shown that a two sublattice structure, caused either by
a lattice distortion, or by the presence of substituents with strong
dipolar nature, can result in negative differential resistance (NDR) in the
transport characteristics. We also find this to be true in case of
correlated insulators. Additionally, we have shown
clear NDR behavior in a correlated double quantum dot by tuning the
electron-electron interaction strength in the system.

\end{abstract}

\section{Overview}

Research in the field of nanoscience has seen an upsurge in
recent times, largely owing to the fact that silicon electronics is fast 
approaching a roadblock, dictated by both the laws of physics as well as the
cost of production. All these years, the semiconductor microelectronics industry has 
always been driven by the need for powerful computational devices 
with high computing speeds. Until now, this has been achieved through 
the `top down' lithographic approach which involves the miniaturization of existing 
silicon-based chips. Gordon Moore, in 1965, in his famous 'Moore's law', predicted that 
there would be a doubling of devices per chip every 18 - 24 months \cite{Moore}, and
this has held true over the past three decades.
However, the `top down' approach is expected to reach its physical limit in the next 
few years owing to certain factors. This rate of downscaling is expected 
to hamper the performance of these devices, as issues related to quantum tunneling, 
interconnect delays, gate oxide reliability, and excessive power dissipation would 
start playing a major role at such small length scales. The electronic properties of 
semiconductor structures fabricated via conventional lithographic processes are 
also quite difficult to control at the nanometer scale. Although some of these issues 
can be overcome by 
improving the device design, the increasing cost of fabrication has 
motivated research in other directions \cite{Ghosh}. It has led to the replacement 
of the `top-down' lithographic approach by a `bottom-up' synthetic chemical approach 
of assembling nanodevices and circuits, directly from their molecular constituents, 
leading to the next generation of electronics, now known as 
`molecular electronics'.

Almost all electronic processes in nature occur in molecular structures,
which are typically of the order of a few nanometer, and hence already possess
a natural scale for use as functional nanodevices. Their abilities of 
selective recognition and self-assembly helps make molecular building blocks through 
cheap methods
of fabrication. Molecules can also exhibit several stable geometric structures or 
isomers, having very different optical and electronic properties. Their 
conformational flexibility can also give rise to interesting transport properties, 
and a simple manipulation of their composition and geometry can lead to a wide variety 
of binding, optical and structural properties, which can be efficiently tuned to 
our needs \cite{Heath,Ayan}.
Moreover, in comparison with molecules, solids have the distinct disadvantage that it is 
relatively difficult and expensive to fabricate them into many millions of nearly 
identical structures, as required in each dense computer chip. 
All these features make molecules ideal candidates
for electronics applications, and there is a growing recognition of this in
the past decade.
\begin{figure}
\centering
\includegraphics[scale=0.7]{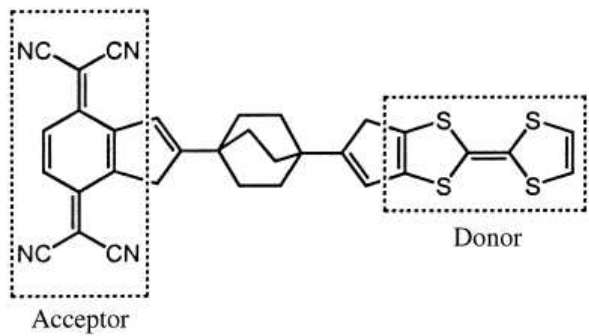}
\caption[Organic analogue of a {\it p-n} junction, composed of a donor 
moiety tetrathiafulvalene (TTF) connected by a methylene bridge to an acceptor moiety 
tetracyanoquinodimethane (TCNQ)]{\label{Aviramfig} Organic analogue of a {\it p-n} junction, 
composed of a donor moiety tetrathiafulvalene (TTF) connected by a methylene bridge to an 
acceptor moiety tetracyanoquinodimethane (TCNQ) (taken from ref. \cite{Ang_review}).}
\end{figure}

The first suggestion, that molecules could indeed be used as alternatives to silicon chips
came from Aviram and Ratner \cite{Aviram} who, in 1974, 
discussed theoretically the possibility to construct a molecular rectifier, based on 
a single organic molecule. They suggested that a single molecule with a 
donor(D)-spacer-acceptor(A) structure (Figure \ref{Aviramfig}) would behave as a p-n junction diode 
when placed between two electrodes. This hypothesis remained so, until more than 20
years later, Metzer \cite{Metzger} studied Langmuir-Blodgett (LB) films of $\gamma$-(n-hexadecyl) 
quinolinium tricyanoquinodimethanide (D-$\pi$ spacer-A species) between metal electrodes and 
demonstrated rectifying behavior. 
Over the years, with the advent of the 
scanning probe microscopy and other techniques, researchers have developed 
ways of addressing, imaging, manipulating, and performing measurements on
molecules connected between metal leads. Advances in synthesis of organic 
molecules, their assembly and measurement has also led to an increasing interest 
in the field of molecular electronics.
Charge transport through molecules can now be probed in a 
controlled way and several prototype devices such as conducting wires, 
rectifiers, switches, and transistors have already been 
demonstrated, as detailed below.

In 1997, Reed et al. measured the conductance of a single molecule of 
benzene-1,4-dithiol in a mechanically controllable break 
junction \cite{Reed_Science_1997}, where a single organic
molecule was adsorbed in an adjustable tunnel gap formed by mechanically
breaking a metal wire on a substrate. 
More complex anthracene based molecules were later
studied by Reichert et al. using the same method and wire and diode like 
properties were demonstrated in symmetric and asymmetric molecules respectively
\cite{Reichert_PRL}. Kushmerick and co-workers used the crossed wire method to measure
transport through symmetric and asymmetric oligo-phenylene-ethynylenes (OPEs),
observing molecular wire and diode features in their respective I-V
characteristics \cite{Kushmerick}. Reed and Tour et al. discovered very interesting 
negative differential resistance (NDR) behavior in these OPEs, functionalized by  NH$_2$ 
and NO$_2$ groups, triggering a gamut of experimental and theoretical studies
to understand the phenomenon, as will be elaborated later.
A related molecular response to an external electric field is that of switching, which is
a very useful property in the design of logic based molecular devices.
In this regard, the most popular class of molecules are catenanes and rotaxanes
which have shown possibilities for being used as switching devices.
Molecules such as bipyridyl-dinitro oligophenylene-ethynylene
dithiol (BPDN-DT) \cite{BPDN} and thiol substituted oligoaniline \cite{Mayer_2005} 
molecules have also been demonstrated to show bistable conductance switching behavior.
pH \cite{Lindsay_NL} and photo induced \cite{photoswitch} molecular switching have 
also been observed in some molecules.
The other most challenging and yet most critical step towards the ultimate 
goal of molecular electronics, is the demonstration of a molecular field-effect 
transistor (FET). Even this has been made possible by 
Xu et al., who have described such a device made of perylene tetracarboxylic 
diimide, where they showed how a variation in the 
electrochemical gate voltage could result in a 1000 fold increase in the source-drain 
current, just like in a n-type FET \cite{Tao_FET}. 

  Single electron transistor (SET) behavior has already
been observed in transport through semiconductor quantum dots, metallic and 
semiconducting nanoparticles, and even in single $\pi$-conjugated organic 
molecules with several distinct charged states which can control its
transport properties.
Carbon nanotubes, fullerenes and semiconductor nanowires represent another
set of interesting systems which have been shown to exhibit Coulomb
Blockade \cite{coul.block1,coul.block2},
Kondo behavior \cite{Kondo,babic.prb} and have also been fabricated into FETs 
\cite{Dekker_nature} and non-volatile memory elements.
Apart from these,  bio molecules, which have both natural self-assembly and self-recognition properties,
have become popular candidates for electronics applications.
A large number of transport measurements have been performed on DNA, the blueprint of life,
which have reported results ranging from conducting, semi-conducting to 
even insulating behavior \cite{Braun,Fink,Porath,Storm,Kasumov}, depending on
experimental methods, surrounding structures, solvents used etc. A protein based 
field-effect transistor (Pro-FET) based on the blue copper protein azurin has also been
demonstrated to operate at room temperature and ambient pressure \cite{PFET}.

\subsection{Theoretical issues}

Although there has been such a quantum increase in the experimental
demonstration of prototype molecular devices, a theoretical understanding
is challenged by a number of fundamental issues. 
A nanoscale molecule between macroscopic electrodes is in a complete state
of non-equilibrium, with each of the source and drain contacts trying
to bring the molecule into equilibrium with its electrochemical
potential, thus driving current through the system.
The problem presents many characteristic length scales, defined to
truly determine when the classical regime ends and the quantum regime begins.
These are the Fermi wavelength, the momentum relaxation
 and phase relaxation lengths, which gain prominence
in this length scale \cite{Tao,SDatta}. Since the system is confined
in one or two directions, the quantum modes in the directions normal to the 
electron propagation direction are discretized, depending on the Fermi wavelength. 
It was shown by Landauer \cite{Landauer}
that the conductance of a nanoscale system depends on 
the transmission probabilities of electrons through these modes, resulting in
quantization of their conductance. A changing width hence results in
steps in the conductance as opposed to continuous dependence
of conductance on dimensions in macroscopic conductors.
The momentum and phase relaxation lengths determine how far the electrons
in the system would travel, before collisions would make them lose
memory of their initial momentum and phase, respectively. While the former
can change the current-voltage response of the device quite drastically due
to impurity scatterings, the latter could take the conduction from a 
incoherent classical transport to a coherent transport where quantum
interference effects would play a very significant role. Classical transport, which
occurs when the length of the conductor is greater than both the momentum
and phase relaxation lengths, resembles transport that we are very familiar
with in the macroscopic regime, obeying the well-known Ohm's law. At the
other extreme, when the system under consideration is smaller than both these length 
scales, transport is known to be ballistic, where the rate of transport
is independent of the length of the system and described by the 
Landauer's formula in terms of transmission probabilities as:

\begin{equation}
G_c = {2e^2 \over h} M T
\end{equation}

\noindent Here, the conductance scales linearly with the transmission ($T$) and the 
number of eigenmodes ($M$) in the wire. Interestingly, the conductance of even a 
ballistic sample is finite, due to the resistance at the interface between the small 
system and the large contacts \cite{Tao}. Measurements on atomic point contacts and 
metallic carbon nanotubes have already demonstrated this behavior where the transmission 
of each available channel is nearly unity. However, in general, the molecular eigenstates
are not fully delocalized leading to a transmission that might be less than unity.
The other regime, where momentum changes occur from impurity scatterings, but with phase 
maintenance, results in a coherent but diffusive transport.
The rate of electron transfer in this case is exponentially dependent on the
length of the molecular bridge. This mechanism holds, especially for short wires with
large HOMO-LUMO gaps such as oligoalkanes. It is also known sometimes as
``superexchange" where the electron transfer proceeds through ``virtual"
orbitals, which are energetically well-separated from the Fermi levels
of the electrodes. In addition to all these, conduction could also be mediated by 
quasiparticles like solitons and polarons in the molecule. 

Apart from length scales that differ markedly when one goes from a macroscopic
to a nanoscale conductor, the energy scales of the whole system consisting of 
the molecule with its discrete energy levels attached to macroscopic metallic 
electrodes with a continuous band structure, poses much
of the challenge in understanding molecular electronic transport.
Some of the issues that one encounters in this regard
are \cite{Zahid} (a) {\it Electrode-molecule coupling and contact surface physics} which 
determine the broadening of the molecular energy levels and hence the lifetime of the 
electron in those levels as well as where the Fermi energy of the electrode
would align with respect to the molecular levels (b) {\it Electronic structure of 
the molecule} which decides the nature of the molecular orbitals and hence their
transmission probabilities (c) { \it Device 
electrostatics} or how the applied external electric field would impact the 
device energetics (d) { \it Inelastic and thermal effects} which come into play when 
strong electron-electron and electron-phonon interactions are present in the molecule.
The most widely employed method for calculating non-equilibrium transport
in such nanoscale systems, with consideration of most of these issues
mentioned  above, is the Non-equilibrium Green's function formalism (NEGF) 
\cite{SDatta}, which for the case of coherent transport boils down to
the Landauer's formalism for calculating current \cite{Mujica_JCP_1994}:

\begin{equation}
\label{Land_eqn}
I={2e \over h}\int_{-\infty}^{\infty} dE \, \, Tr[\Gamma_LG\Gamma_RG\dag] \; \; (f(E,\mu_L)-f(E,\mu_R))
\end{equation}

\noindent Here $G$ represents the device Green's function, $\Gamma_{L,R}$, the 
imaginary part of the self-energies correspond to the broadening of the molecular 
energies, and f(E) represents the Fermi-Dirac distributions at 
the two electrodes with electrochemical potentials $\mu_L$ and $\mu_R$.. 
The NEGF formalism has been combined with many semi-empirical and Hartree Fock
methods to obtain the current-voltage characteristics of molecules. Transport 
calculations have also been incorporated into electronic structure methods and 
packages such as TRANSIESTA \cite{Transiesta}, Atomistix toolkit (ATK), 
WanT \cite{WanT}, McDCal \cite{McDCal}, Gaussian 
embedded cluster method \cite{GECM} etc, which have become very popular in present times.

In this article, we discuss various aspects of a molecule connected between macroscopic
electrodes. In section II, we briefly discuss the effect 
of external electric field on the potential profile in an interacting 
molecular wire. In section III, we introduce
the phenomenon of negative differential resistance in single molecular junctions,
and detail the experimental and theoretical efforts over the 
last few years in understanding the phenomenon. We will then elaborate
our work on this interesting device behavior, focusing on the effect of
dimerization and donor-acceptor groups in causing NDR. The insight that we have gained 
from our work on NDR in molecular systems (which are primarily insulating due to their
finite sizes), resulted in studies on the effect of electric
field on insulators formed out of the strong
electron-electron interactions in them (such as the Mott insulators). This part
will be briefly described in section IV. In section V we give a detail
description of NDR in the coulomb blockade regime in molecular quantum dots.
Finally, we conclude with the summary of results and discuss the future outlook.

\section{The potential profile in an interacting \\ molecular wire}

As was discussed before, a variety of factors have been
found to be very crucial in determining the nature of current-voltage
characteristics in nanoscale systems, like the electronic
structure of the constituent molecule, the interface physics and the
profile of the potential drop across the molecule between the electrodes.
This latter point, in itself, is very interesting to probe, as most often,
it is not the low bias or linear response characteristics of a wire
that are interesting, but rather the full current-voltage characteristics.
The low-bias response would be determined simply by the equilibrium
electronic structure and energies, but the device characteristics
at larger voltages would require one to understand in detail the nature of the 
potential profile inside the molecule in response to an externally applied 
voltage.

\begin{figure}
\centering
\includegraphics[width=0.9\columnwidth]{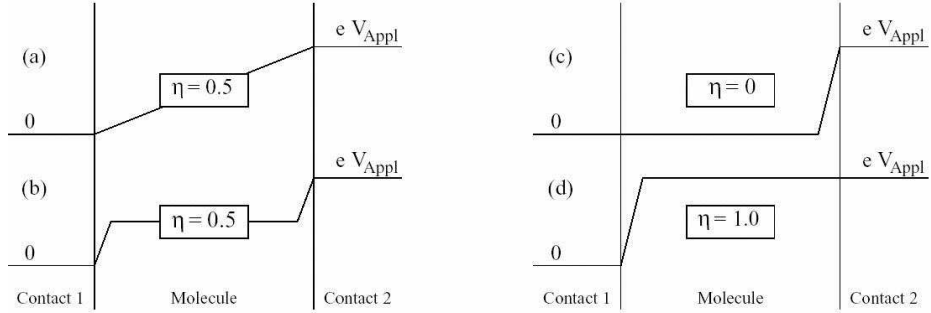}
\caption[Examples of a few potential profiles in terms of the
parameter $\eta$]{ \label{etafig} Examples of a few potential profiles in 
terms of the parameter $\eta$} (taken from reference \cite{Zahid}).
\end{figure}
Datta et al. \cite{Zahid,Datta_PRL_1997}
suggested a one parameter characterization of this potential profile, through the
voltage division factor $\eta$ ($0<\eta<1$), which rigidly changes the molecular energies
by $\eta e V$ or equivalently, shifts the electrochemical potential of the left and right 
electrodes as $\mu_L=E_F-\eta e V$ and $\mu_R=E_F+(1-\eta) e V$. Figure \ref{etafig} shows
few examples of the potential profile obtained by varying the values of $\eta$. 
They found that this parameter has a profound effect on the molecular current-voltage 
characteristics. A value of $\eta=0.5$ occurs when the molecule is coupled equally to both 
electrodes, however as shown in Figure \ref{etafig} (a) and (b), it could still result in 
two different potential profiles depending on the extent of charging in the molecule. 
The current through the molecule, in this case, especially at low bias, 
would be nearly symmetric. Values of $\eta=0$ and $1$  
appear when the coupling to the contacts are unequal. This would
result in a I-V curve whose positive branch would look very different from 
the negative one.

Most of the earlier calculations in this field assumed the electrostatic 
potential profile due to the applied bias to be a ramp (linear) function.
This is equivalent to assuming that all the charge is localized only in 
the external boundaries, neglecting any effects due to screening.
However, at high bias, there could be a temporary accumulation of charge on 
the bridge, which would screen the external field and hence result in
a profile with non-uniform field gradient. To account for this, Mujica et al. have
proposed a self-consistent solution of Poisson equation coupled with
the Schr\"{o}dinger equation \cite{Mujica_JCP_2000} and suggested that this 
is essential to account for the actual features of
the current spectra of these nanowires. In our studies \cite{lakshmi_Proc}, 
we have included this 
aspect while recognizing the effects of electron-electron interactions
on the potential profile across a molecular wire, assumed 
to be a one-dimensional chain of $N$ atoms with one orbital per atomic site.
We assume a Hubbard model to describe the system as
\beq
\label{exacteqn} H =  \sum_i\sum_\sigma -t(a^{\dag}_{i\sigma}a_{i+1\sigma}+hc)
  +  U\sum_i n_{i\uparrow}n_{i\downarrow}
\eeq
\noindent where $t$ is the electron hopping term and $U$ is the Hubbard 
on-site electron repulsion term. To solve this Hamiltonian in the 
mean-field limit with electronic spins explicitly, we consider the 
averaged mean-field quantities, $\langle a^{\dag}_{i\sigma}
a_{i\sigma}\rangle$ and 
$\langle a^{\dag}_{i\sigma}a_{i-\sigma}\rangle$ \cite{SKPati_JCP,Fradkin}.
The averaged (mean-field) form of the Hamiltonian can then be written in terms 
of these mean-field quantities as, 
\begin{figure}
\centering
\includegraphics[width=0.7\columnwidth]{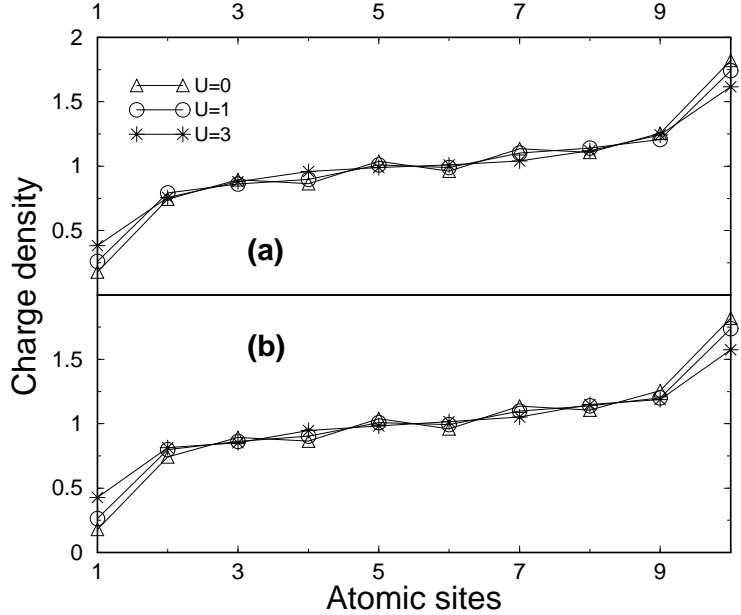}
\caption{ \label{cdenfig} Comparisons of charge densities as a function of atomic sites 
for various $U$ parameters with mean-field (a), and exact (b) methods, calculated at a 
bias of 6V.}
\end{figure}

\beq
H_{mf} =  -t\sum_i(a^{\dag}_{i\sigma}a_{i+1\sigma}+hc) 
+ U\sum_{i\sigma}a^{\dag}_{i-\sigma}a_{i-\sigma}\langle a^{\dag}_{i\sigma}
a_{i\sigma}\rangle \nonumber \\
 -  U\sum_i(\langle a^{\dag}_{i\sigma}a_{i-\sigma}\rangle 
a^{\dag}_{i\uparrow}a_{i\downarrow}+hc). 
\eeq

\noindent Note that, the second $U$ term in eqn. (4) destroyes the spin 
conservation and would only contribute if the system has a nonzero finite 
magnetization. In our case, however, we consider the chain of $N$ atoms with 
$N/2$ up spins and $N/2$ down spins, so that the ground state magnetization is 
zero. Therefore, although this term does not contribute for our system, the
term is explicitly given since the Hamiltonian describes the general mean-field
derivation of the Hubbard model.

 The electric field applied on the wire adds the term 
$\sum_{i,\sigma} F_i a^{\dag}_{i\sigma}a_{i\sigma}$ to the Hamiltonian, where 
the variation of the field $F$ with the site index $i$
is the potential profile we are interested in. We start our calculations by assuming
that the electrostatic field is a linear ramp function across the metal-molecule
interface. By solving the mean-field Hamiltonian, we obtain the charge density
($\rho_i$) at every site, given by  
$\rho_i=\sum_\sigma a^{\dag}_{i\sigma}a_{i\sigma}$. This becomes the input
for the one-dimensional Poisson equation,
\beq
F_{i+1} + F_{i-1} -2 F_i=-\rho_i
\eeq

\noindent where the inter-atomic distance and the dielectric constants at 
every atomic site are assumed to be constant and unity. 
Additionally, unlike previous self-consistent studies \cite{Mujica_JCP_2000}, 
where the boundary conditions for solving the Poisson equation was considered 
by assuming finite but large dielectric constants for the metal electrodes as compared to the 
molecular sites, here we enforce that the left electrode (to which the atomic site ``1" is 
weakly coupled) has zero bias while the right electrode has the full bias (ie, $F$). 
This ensures that the electrodes are not polarized due to charge accumulations at the 
molecular sites. We then solve the modified Hamiltonian
$\tilde{H_{ii}}=H_{ii}-eF_i$, self-consistently with the Poisson equations 
using the appropriate boundary conditions, until the charge densities and the site potential 
fields converge.

Since our
calculations are performed in the mean-field, we have also compared our results 
with those obtained by performing exact diagonalization calculations.
Our charge density calculations on a 10 site half-filled system, with various values 
of the Hubbard strength $U$, shows uniform density in the middle of the chain
(with alternations in the ends due to boundary effects), at zero bias.
However, with increasing bias, as shown in Figure \ref{cdenfig}, a depletion of charge 
densities in one end, and an accumulation in the other, ensues. Furthermore, 
the induced polarization due to the external field decreases
with increase in $U$. This is quite easy to understand, since, increase in $U$ 
localizes the charges, and so the effective field that the system experiences is 
smaller for larger $U$ values. It can also be seen that for all values of $U$, 
the mean-field charge densities compare fairly well with the exact calculations.

\begin{figure}
\centering
\includegraphics[width=0.6\columnwidth,angle=-90]{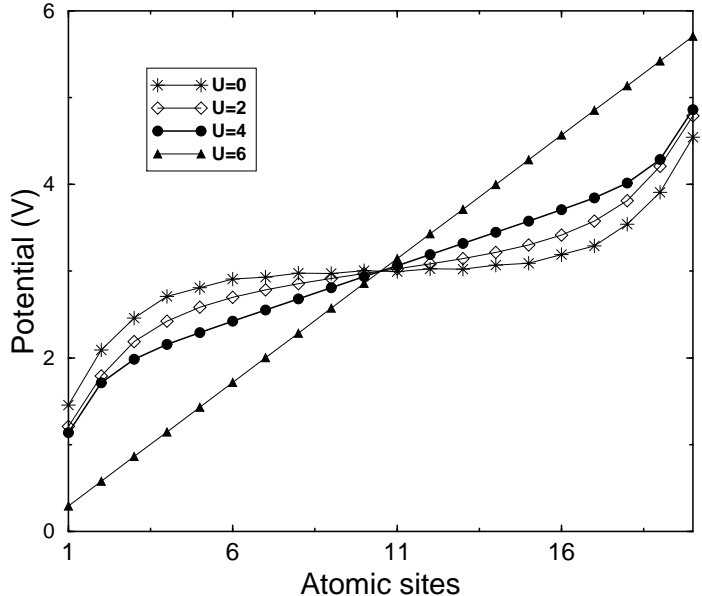}
\caption{\label{profilefig1} Spatial electrostatic potential profile across the
electrode-molecule interface, for various $U$ values,
calculated at a bias of 6V.}
\end{figure}
In Figure \ref{profilefig1}, we present the self-consistently derived potential profile 
for a chain of 20 sites, with different $U$.
At the outset, our calculations show that for a tight-binding molecule ($U=0$), the potential
has the features which has been described by others \cite{Mujica_JCP_2000}, 
in which most of the voltage drop occurs near the electrodes,
and is essentially constant (zero gradient) in the wire region (which 
we will call the tight binding potential 
profile or TBPF) . The tight-binding solutions are plane-waves wherein 
electrons are completely delocalized all over the molecule. The inclusion of
a finite Hubbard repulsion is known to induce Mott localization, and the
potential differs from the tight-binding solution, with the gradients becoming 
nonzero in the middle of the molecule. The general feature, hence shows 
that, increasing the value of $U$ from small to large takes 
the potential profile slowly from TBPF to ramp spatial variation.
This is because large $U$ introduces strong site
localization and the electrons are essentially ``particle-like" in the large
$U$ limit. Hence the external field is not able to induce much accumulation
of excess charge in the system, thus retaining the ramp nature of the potential
profile.

These studies give an idea of how interactions in a nanoscale system can
change the nature of the potential profile across it, which is expected to
have a lot of implications in its current-voltage characteristics also.

\section{Negative Differential Resistance}

\begin{figure}
\centering
\includegraphics[scale=0.7]{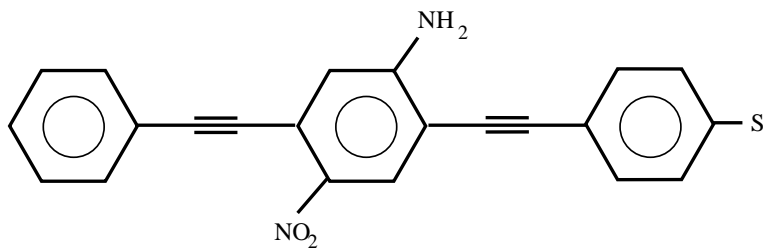}
\caption{\label{molecule} The Tour molecule consisting of three benzene rings 
connected by triple bonds with donor (NH$_2$) and acceptor (NO$_2$) substituents 
in the middle ring.}
\end{figure}

Negative differential resistance (NDR) behavior, is a property of electrical circuits 
in which, over certain voltage ranges, current is a decreasing function of voltage. 
Semiconductor devices exploiting this effect are widely used to make 
amplifiers and oscillators, however, it was reported for the first time in
molecular systems, by Reed and 
Tour et al. in oligo(phenylene-ethynylenes) 
(OPEs), functionalized by amine (NH$_2$) and nitro (NO$_2$) substituents 
\cite{Chen_science_1999}.
This molecule shown in Figure \ref{molecule}, 
consists of three benzene rings connected by triple bonds, with donor (NH$_2$) 
and acceptor (NO$_2$) groups (or sometimes only the
acceptor) in the middle ring. The strong NDR peak with a 
peak-to-valley ratio (PVR) of 1030:1 was observed at 60K, and was found 
to reduce with increasing temperature \cite{Chen_APL_2000}.
They also reported electronically programmable memory behavior in the NO$_2$
functionalized devices \cite{Chen_RAM}, using the 
nanopore setup for performing experiments. Very recent studies performed 
by Tao et al., using a STM break 
junction method where a Au tip was driven into contact with a Au surface in a 
solution containing the organic molecules have proven to be very effective 
{\it single} molecule junctions. The conductance was then measured as the tip 
was retracted. Their experiments also confirmed earlier observations by 
indicating sharp NDR peaks in the NO$_2$ functionalized OPEs \cite{OPE_Tao}. 
NDR response has also been reported in 
ferrocenylundecanethiol self-assembled monolayers \cite{FeSAM}, azobenzene
molecular junctions \cite{NAB}, oligopeptides \cite{Opep}, C$_{60}$ encapsulated
double walled carbon nanotubes \cite{C60_DWNT}, bilayer junctions formed by nitro
substituted OPE and an alkanethiol \cite{RAKiehl}, and even in molecules 
on doped silicon surfaces \cite{NDR_Hersam}.

A large number of theoretical studies has also been performed
on many molecules to understand the underlying causes for molecular NDR 
behavior, sometimes leading to even predictions of
NDR in some novel molecules. Theoretical predictions of NDR on systems include 
carbon atomic wires \cite{Atomicwires},  
metallic and semiconducting clusters \cite{MSC}, molecules between transition metal 
contacts \cite{TrCon} and some organometallic molecules \cite{Baranger}.
Considerable amount of theoretical work based on semi-empirical and
{\it ab}-initio methods have also been performed to understand 
NDR behavior in the Tour molecules.  Many explanations for this phenomenon based on 
charging \cite{Seminario_JACS_2000,Seminario_JCP_2002,2site}, reduction of the acceptor 
moiety \cite{Xiao_JACS_2005}, twisting of the ring structure leading to conformational 
changes \cite{Taylor,Bredas_JACS_2001,Cornil_JACS_2002,RPati_PRB_2004}, bias driven 
changes in molecule-electrode coupling \cite{XShi_JPCB_2005} etc, have been proposed. 
However, most of these require to impose some external factors like the rotation of the 
middle ring, or introduction of extra charge in the molecule in order for the 
external bias to cause NDR at some bias. Also, most often they do not make a 
relation between structural preference and bias polarity and hence do not 
explain the asymmetry that has been observed in the experimental I-V characteristics. 
In our studies on this phenomenon, we have used simple, but insightful models to
understand the NDR behavior. Our study involves two aspects of these molecules
(a) conjugation or dimerization (b) presence of donor-acceptor substituents and their
respective roles in taking the molecule through a negative differential
resistance, when subjected to an external electric field.

\subsection{Role of dimerization (lattice distortion)}

It is well-known that a half-filled one-dimensional metal,
due to the effects of strong electron-phonon coupling, is unstable against 
Peierls distortion, producing a finite charge gap in the system, thereby 
lowering the energy of the occupied states and stabilizing the 
distortion \cite{peierls}. The competition between
the lowering of the electronic energy and the increase of 
the elastic energy due to distortions often
leads to modulation of the bond lengths in the system, removing 
high density of states at the Fermi surface.

For such low-dimensional systems, the approximation that the electron 
transfer occurs through purely the electronic states, completely ignoring 
the change in underlying lattice structure during the transfer, is not valid. 
There have been many studies on the inelastic conductance through molecular
wires in the presence of vibrational modes by Ness, Fisher \cite{Ness_Fisher}, 
May, Todorov, Segal, Nitzan \cite{GME} and 
others before. However, in our study, the assumption of strong coupling between the molecule 
and electrodes suggests that the electron traversal time is large and hence we neglect 
possibilities of dephasing due to electron-phonon interaction. The time scale is also too short 
for the electron to remain on the molecular bridge and hence form a polaron.
However, the net effect of the interaction is to distort the molecular bridge resulting in
modified electron hopping strengths. 
In this study \cite{lakshmi_JCP}, hence, we have considered transport properties 
purely in the presence of underlying lattice distortions.
We describe the finite wire within the Su-Schreiffer-Heager (SSH) model, where the 
electrons are treated using a tight-binding description and the lattice degrees of freedom 
are treated adiabatically\cite{ssh1,ssh2}. The interaction between the wire and electrode are 
considered through the Newns-Anderson chemisorption model 
\cite{Newns_PR_1969,Anderson_PR_1961} and the nonlinear response of the current 
with the applied bias calculated using the Landauer's formalism.
The Hamiltonian is given by:
\bear
H & = & \sum_i-(t+\alpha(u_i-u_{i+1}))(a^{\dag}_{i}a_{i+1}+hc)
+{1\over {2}}K\sum_i(u_{i+1}-u_i)^2
\eear

\noindent where the first term describes the hopping of $\pi$ electrons
along the polyene chain without spin flip, the second term 
corresponds to
the $\pi$ electron-phonon interaction, and the last term is the
phonon Hamiltonian. $t$ is the nearest neighbor hopping integral
and $K$ and $\alpha$ are the elastic spring constant associated with
the nuclear motion and the electron-phonon coupling strength
respectively. In this
Hamiltonian, the hopping strengths are linearized, i.e, the  
strength of the hopping is proportional to the difference in
displacements $u_i$ of the atoms from their mean positions.
The Hamiltonian can be re-written as
$H  =  \sum_i-(1+\delta_i)(a^{\dag}_{i}a_{i+1}+hc)
+{1\over {\pi\lambda}}\sum_i\delta_i^2$
 where $\delta_i = {\alpha(u_i-u_{i+1})\over{t}}$ and
$\lambda$ is a dimensionless coupling constant
defined as $\lambda = \frac{2\alpha^2}{\pi K t^2}$, which we
will use in our calculations as a measure of the electron-phonon
coupling strength. We minimize the above Hamiltonian with the constraint that 
the energy changes associated with
the net displacement about the mean positions should be zero so
that the total length of the wire remains constant, i.e, 
$\sum_i\delta_i=0$. 
Minimizing it with respect to the $\delta_i$
gives an expression for the $\delta$s, the energy shift of the i$^{th}$ bond from
its equilibrium value,
\beq
\delta_i=\frac{\pi\lambda}{2}[ \langle a^{\dag}_ia_{i+1}
+hc \rangle - \frac{1}{N} \sum_i \langle a^{\dag}_ia_{i+1}+hc \rangle ]
\eeq
\noindent where $N$ is the total number of bonds. 
We have calculated these $\delta$s, first, for 
an isolated wire consisting of 20 sites for a range of electron-phonon  
coupling strengths. We find that as the coupling strength is increased, the 
chain is more and more distorted (dimerized), however, in this process, the 
system gains some additional energy, and opens
up a gap about the Fermi energy thereby making the system insulating.
This is the Peierls dimerization mechanism.
This nonconducting gap increases as the coupling strength
is increased. For $\lambda=1$, we find that the HOMO-LUMO gap is 
about $1$eV, a value typically observed for a long but finite polyene system,
and which we will use for further calculations. 

\begin{figure}
\centering
\includegraphics[scale=0.5]{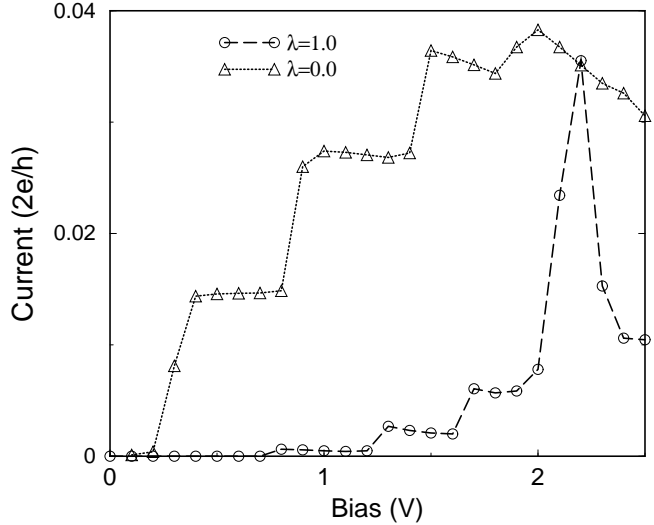}
\caption{ \label{fig3_even} The current-voltage characteristics for a 20 sites 
molecular wire for the non-interacting case ($\lambda=0$, triangles) and with
electron-phonon coupling ($\lambda=1$, circles).}
\end{figure}
As a first step to understanding device response, we look at the
equilibrium transmission through the states around the Fermi energy($E_f$).
Quite interestingly, it shows that unlike in a pure non-interacting
wire, the presence of electron-phonon
coupling localizes the orbitals close to the Fermi energy, even
in the absence of external bias. These are orbitals which are very
important from transport point of view and we will show later, the
consequences of such a localization.
We next calculate the current-voltage (I-V) characteristics of the
20 sites molecular chain for $\lambda=1$ 
as shown in Figure \ref{fig3_even}. 
As can be seen, the current increases by jumps for
the noninteracting chain, and this staircase structure is quite 
well understood \cite{Zahid, Mujica_JCP_1996}. 
However, with the inclusion of electron-phonon coupling, the I-V curve 
shows some remarkable features. Plateau structure is no more present at high
bias and we find a sharply peaked NDR structure in the off-resonant
condition. Interestingly, the magnitude of current
at the critical bias is same with and without the electron-phonon 
coupling. 

To trace back the origin of this NDR behavior, we
look at the variation of the low-energy levels (HOMO and LUMO) 
with bias, and find that in general the 
levels stabilize linearly with bias at small bias
strength. But at the bias at which NDR is seen (critical bias), 
the HOMO and LUMO levels come very close up to a gap of 
$\sim 0.05$eV, after which they move farther apart.
In the presence of strong electron-phonon coupling this 
accidental degeneracy between
the levels induces mixing between the states that would be 
absent otherwise. Calculating other quantities such as the
numerator of the Greens' function (which determines the nature of the
plateau structure in the I-V) and the inverse participation
ratio (IPR), we confirm that the NDR occurs when the system  goes
from one localized phase to another through a completely delocalized 
phase, at the critical bias, resulting in a sharp rise and fall
in current.

We now explain this localization versus delocalization features 
of the levels as a function of bias with electron-phonon coupling.
In the absence of electron-phonon coupling, all the bond lengths of the
system are equal. The ground state wavefunction, corresponding to the HOMO 
level, consists of alternate symmetric and antisymmetric combinations of
the atomic coefficients for the successive bonds across the chain. This corresponds 
to a particular parity configuration. However, since the molecular levels 
are placed in energy with alternate parity corresponding to the 
alternate energy levels, the LUMO will have 
exactly the opposite parity to the HOMO level \cite{Keiss}. The wavefunction 
for LUMO
will then consist of opposite combinations of the atomic coefficients for
the corresponding bonds.
If the HOMO is written as $|++--++-->$, the LUMO then corresponds to
$|+--++--++-->$, where the $+(-)$ corresponds to the sign of the atomic 
coefficients for the bond. 
However, as has been discussed, the system becomes dimerized with alternating
long and short bonds once the electron-phonon coupling is introduced.
The symmetric and antisymmetric combinations of the atomic
coefficients in this case now correspond to the long and short bonds
(or short and long bonds) respectively. In other words, if the HOMO is 
mapped to a lattice configuration with bond length variation as $=-=-=-$, 
the LUMO will then be $-=-=-=$. 
\begin{figure}
\centering
\includegraphics[scale=0.6]{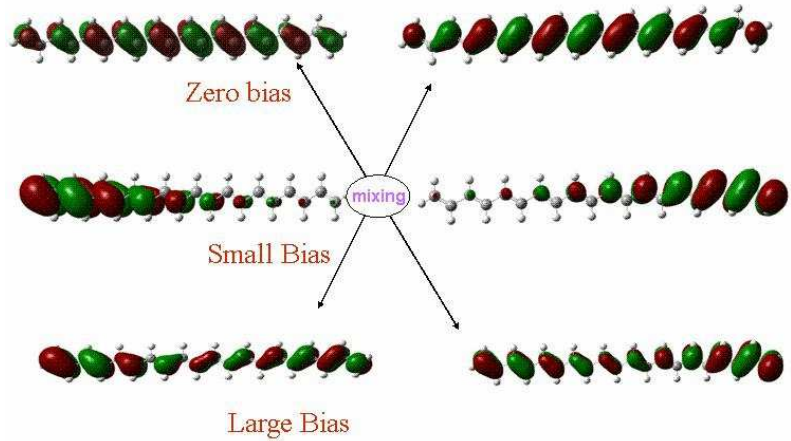}
\caption{ \label{Orbitals} The HOMO (left) and LUMO (right) of the poly-acetylene
chain of 20 atoms, in the presence of no, small and large electric field strengths. }
\end{figure}
The large HOMO-LUMO gap at the zero bias condition 
is precisely the energy difference associated with the dimerized chain 
corresponding to these opposite symmetry combinations. 
This is also very clear from Figure \ref{Orbitals}, where we show
the HOMO and LUMO orbitals for the inherently dimerized
trans-polyacetylene system in the presence and absence of
electric field. The system was optimized
within the Gaussian 03 suite of programs, 
using the hybrid Becke 3 Lee-Yang-Parr (B3LYP) gradient corrected
exchange-correlation functional, at the 6-31G(d) basis set level.
At zero bias, it is very evident that the HOMO is localized on the double bonds 
and the LUMO on the single bonds.
As the bias is applied, the LUMO approaches the coefficient 
combinations corresponding to the HOMO and similarly the HOMO
towards the LUMO, and at some critical bias, both of them
have their energies corresponding to the lattice with equal bond lengths.
Since for equal bond-lengths, the energy difference between these
combinations is quite small,
the HOMO and LUMO at the critical bias are almost degenerate.
A glance at the orbitals in Figure \ref{Orbitals}, at small field, shows
that the HOMO is getting localized in one half of the chain and the
LUMO in the other half.
At the critical bias, the quasi degeneracy of their energies,
results in orbital mixing, leading to a complete delocalization
over the entire chain. A complete confirmation of this is obtained
from the I-V characteristics, where the current at the critical bias with 
electron-phonon coupling is exactly the same as that for the noninteracting
system with equal bond-lengths (see Figure \ref{fig3_even}). With a further increase 
in field, it becomes very clear from Figure \ref{Orbitals} that the HOMO orbitals are 
beginning to look like the LUMO and vice-versa, indicating that there is a change 
occurring from one insulating phase to the other, leading to the fall in current seen in the
I-V characteristics.

\begin{figure}
\centering
\includegraphics[scale=0.56]{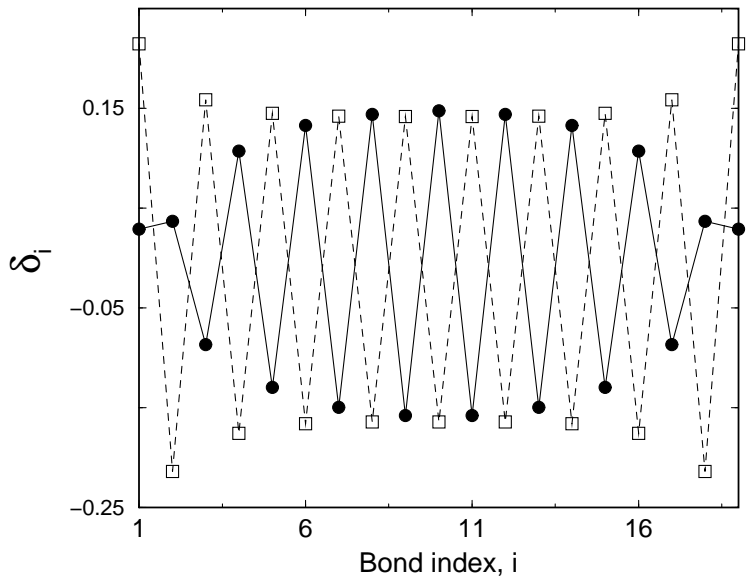}
\caption{ \label{fig6_even} The bond-alternation parameter ($\delta_i$) as a function
of bond index, $i$, as obtained from the Poisson-Schr\"{o}dinger
self-consistent calculations for two values of the bias. Before NDR 
(bias=$2.3$V, squares) and after NDR (bias=$2.4$V, circles)[see Figure \ref{fig3_even}].}
\end{figure}

We have also carried out self-consistent calculations
combining the one-dimensional Poisson and Schr\"{o}dinger equations, as 
described in the previous section. This is to explore the effect of the spatial 
electrostatic potential profile on the bond distortions and the 
transport across the chain. 
Since, in these calculations, 
the distortion pattern ($\delta$s) are self-consistently calculated at 
every bias, we find that, this even site half-filled system goes from 
one kind of dimerized phase to the other around this critical bias at 
which the energy levels show quasi-degenerate behavior. This is shown
in Figure \ref{fig6_even}, where we have plotted distortion patterns at the bias values
before and after the NDR peak in the I-V curve. Since the chain is an open 
system, the opposite dimerization pattern is evident only in the middle 
of the chain. 
At the critical bias, the nature of the potential profile also changes 
from RAMP, which is the profile we start with, to 
the tight binding potential profile (TBPF), which has large
gradients towards the ends of the chains and zero gradient at the middle 
\cite{Mujica_JCP_2000}.
A similar mechanism for NDR was observed in an odd-sites system as well,
where the notable point is that the ground state is a solitonic
state. We find that the SOMO (singly occupied molecular orbital) also 
contributes to the mixing of energy levels at the critical bias,
leading to NDR peaks \cite{lakshmi_Pramana}.

To have a microscopic understanding of this NDR phenomenon, we also derive analytic
expressions for the low-lying energies and wavefunctions, using time-independent 
perturbation theory, where the external bias acts as the perturbation.
We consider two cases for our study - that of a uniform ($\delta t=0$) chain and 
a partially dimerized chain, with the Hamiltonian, 
$H=-\sum_{i,\sigma}(t + (-1)^{i+1} \delta t) (a^\dag_{i,\sigma}a_{i+1,\sigma} + hc)$,
and derived expressions for the first and second order change in energies, caused
by the presence of the field.
The main point is that, in a dimer, the two sites contribute differently
to the ground state of the partially dimerized chain, and thus
gets intermixed by the external field creating a dipolar contribution that essentially 
allows the low-energy levels to approach each other. This leads to possibilities
of NDR in the I-V characteristics of the dimerized chain as opposed to a uniform one.
We refer the readers to Ref.\cite{lakshmi_JPCM} for more details of this study.

\subsection{Role of donor-acceptor groups}
Although our previous study throws
some light on why NDR appears in conjugated systems, this subsection complements
it by investigating the effect of the donor-acceptor groups
in causing this. We find that not only NDR, but also the asymmetry 
that characterizes the I-V characteristics of these molecules,
comes out naturally of our model \cite{lakshmi_PRB}.
And the simplicity of our parametrized model makes it very tractable and 
lends physical insight into the factors causing NDR.

Our model assumes that the part of the Tour molecule with the 
donor group (NH$_2$) has a negative on-site energy and that with the acceptor group (NO$_2$) 
has a positive on-site energy 
The spatial variation of bias on the structure,
is considered (except in few cases considered later) to drop as a ramp 
function, varying linearly from one electrode to the other. 
With this potential, the energies for the dimer with Hamiltonian 
$H  =  \displaystyle \sum_{i=1,2} \epsilon_i(a^{\dag}_{i}a_{i}) + \sum_i -t(a^{\dag}_{i}a_{i+1}+hc)$,
where $\epsilon_i$ is the on-site energy of site $i$ and $t$ is the
hopping integral between the donor and acceptor, can be easily derived as:
\beq
\label{eqnener}
E_{\pm}=\frac{\epsilon_1+\epsilon_2-V}{2} 
 \mp\frac{\sqrt{9(\epsilon_1-\epsilon_2)^2+36t^2+V^2+6V(\epsilon_1-\epsilon_2)}}{6}.
\eeq

\noindent The coupling to the electrodes modifies the bare Greens 
function of the molecule, which can be written as

\bear
G_{12}&(E,V)&=\frac{V-3(\epsilon_2-\epsilon_1)-\sqrt{(3(\epsilon_2-\epsilon_1)-V)^2+36t^2}}{6t (E-E_1+i\Sigma_L+i\Sigma_R)} \nonumber \\
  &+&\frac{V-3(\epsilon_2-\epsilon_1)+\sqrt{(3(\epsilon_2-\epsilon_1)-V)^2+36t^2}}{6t (E-E_2+i\Sigma_L+i\Sigma_R)}
\eear

\noindent where $\Sigma_L$ and $\Sigma_R$ are the self-energies corresponding
to the left and right electrodes. We calculate these quantities within the Newns-Anderson
model, and subsequently current from the Landauer's formula.

\begin{figure}
\centering
\includegraphics[scale=0.45]{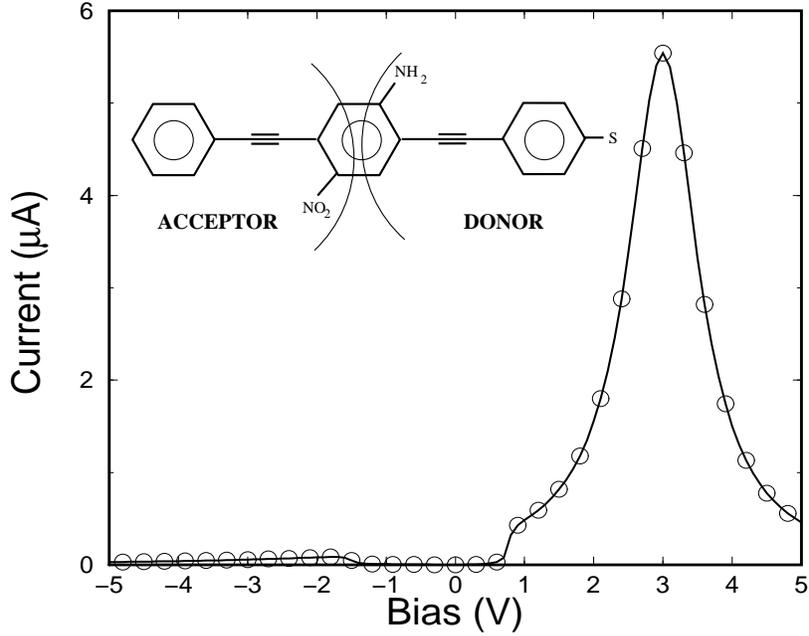}
\caption{ \label{Tour_IV} The current-voltage characteristics for the 2-site 
system for $\epsilon_2=-\epsilon_1=0.5 eV$ and $t=0.1 eV$. Inset is the 
Tour molecule.}
\end{figure}
At zero bias, as can be seen from Eqn.(\ref{eqnener}), with $V=0$, presence 
of different on-site energies opens up a gap larger than that
for a purely hopping model ($\epsilon_2=\epsilon_1=0 eV$)
near the zero of energy indicating the preference of the electrons to stay 
at the atomic site with negative on-site energy. 
The equilibrium transmission is found to be large for purely hopping
model since it corresponds to equal distribution of charges.  
With the inclusion of different on-site energies, the system becomes
insulating due to charge transfer and the zero-bias transmission reduces due to this
preferential charge localization,

Figure \ref{Tour_IV} shows the nature of the current-voltage characteristics with
external bias. As can be seen, the
current is negligible around the zero of energy and around a bias of $1 V$, 
there is a small jump in the current. 
With increase in the forward bias, around a bias of $3 V$, the current shows a 
sharp rise and fall, indicating strong Negative Differential Resistance (NDR). On 
the other hand, with increase in the reverse bias, the system continues to 
remain insulating with negligible current.

\begin{figure}
\centering
\includegraphics[scale=0.4]{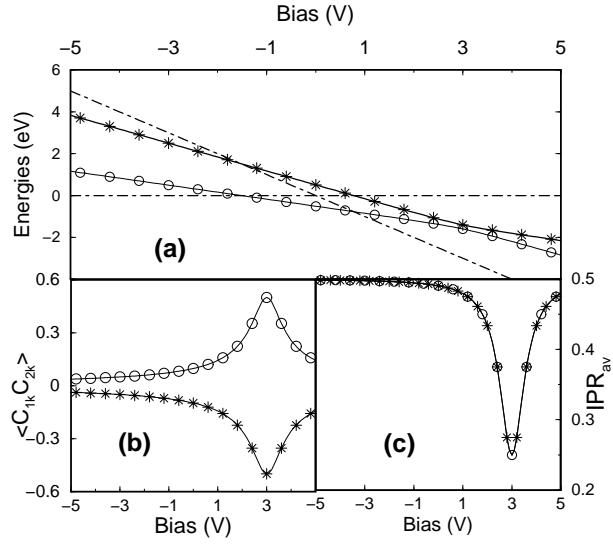}
\caption{\label{ener_IPR} (a) The variation of the two levels (circles and stars) 
of the 2-site system with the applied bias, for $\epsilon_2=-\epsilon_1=0.5 eV, t=0.1 eV$
The dotted lines indicate the variation
of the Fermi energies of the electrodes with bias.
(b) The numerator of the Greens function matrix element for the corresponding
energy levels shown in (a). 1 and 2 represent the site index and k specifies
the corresponding level.(c) The $IPR_{av}$ for the levels shown in (a). }
\end{figure}

To understand the reasons for the NDR, we look at the variation 
of the energy levels ($E_k$) of the bare molecular dimer with bias 
(Figure \ref{ener_IPR}(a)) and the numerator of the Greens function, 
$\langle 1|k \rangle \langle 2|k \rangle$ where $k=1,2$ are the eigenstates
(Figure \ref{ener_IPR}(b)), as a function of the external bias. With increase 
in the forward bias, the energy levels come close to one another up to
the critical bias $V_{c}$ at which the NDR is seen, after which, they 
move farther away. In Figure \ref{ener_IPR}(b), exactly around this $V_{c}$, the
contribution to the dimer eigenstate coefficients from the sites increases quite sharply,
indicating a more delocalized state.
We also calculate the average inverse 
participation ratio ($IPR_{av}$) which defines the extent of localization for 
a given eigenstate, with energy $E_k$:
\beq
IPR_{av} = \frac{1}{D(E)}\frac{1}{N}\sum_{k} P_k^{-1}\delta(E-E_k)
\eeq

\noindent where $P_k^{-1}$ is the IPR, defined as 
$P_k^{-1}=\frac{1}{N}\sum_{j}|{\psi(j,k)}|^4$ where the $j$ is
the atomic site index and $D(E)$ is the density of states. 
Figure \ref{ener_IPR}(c) shows a strong dip in the values of $IPR_{av}$
around the critical bias confirming complete delocalization in the system, while
at other values of the bias, $IPR_{av}$ is much larger due to the localized
nature of the eigenstates.

Since the model is exactly solvable, we quantify this critical bias $V_{c}$,
by minimizing the gap between the energies with respect to the applied bias and obtain,
$V_{c}=3(\epsilon_2-\epsilon_1)$, which is in accordance with our numerical data
(for $\epsilon_1=-0.5 eV$ and $\epsilon_2=0.5 eV$, we find $V_{c} \sim 3 V$) \cite{Pearson}.
[Using the values of the absolute electronegativities and hardness, the ionization 
potential difference between amino and nitro group is calculated to be of the
order of 1.0 eV, same as what we considered for $\epsilon_2-\epsilon_1$. 
Critical bias calculated from this, gives a reasonable comparison with
experiments].
At this critical bias, the energies take the values $\mp t$, precisely the
energies of the system with $\epsilon_1$=$\epsilon_2$=0.
However, with increase in the reverse bias, the energy levels start 
diverging away from their
zero bias gap making the system more insulating, explaining 
the small current that is observed in Figure \ref{Tour_IV} for negative bias.

We argue below physically what happens to the dimer in the presence of
both the forward and reverse bias.
Initially for small bias, as noted before, the system tends to accumulate 
its charge density at the site with lower
on-site energy. Such a localization makes the system insulating. If this site 
is closer to the electrode with higher chemical potential, an increase in
bias makes the charges tend to move 
towards the other site. When the bias equals the critical bias $V_c$, where the 
NDR is seen, the charge densities are equally distributed at both
sites with no preference of one site over another, describing a situation
where both the on-site energies are equal. Further increase of bias would 
localize the charges
on the other site resembling an insulating dimer with its on-site 
energies interchanged, precisely the case as in the reverse bias situation.


\begin{figure}
\centering
\includegraphics[scale=0.45]{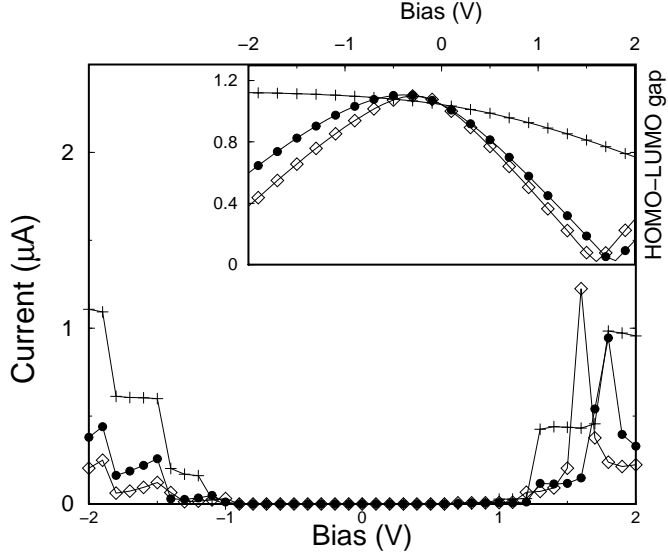}
\caption{ \label{IV_profile} The I-V characteristics for the 20 site system with 
$\epsilon_2=-\epsilon_1=0.5 eV$ and $t=1 eV$ with ramp potential 
 (diamonds), l=2 drop (plus), and l=8 drop (circles) close to the
interface. The inset shows the variation of the HOMO-LUMO gap
with bias for the three cases.}
\end{figure}
Extending the dimer model to a $N$ sites chain with alternating donor and
acceptor sites, we can, from a perturbation view-point, derive an expression 
for the critical bias for NDR to be $V_c|_{t \rightarrow 0}=(\epsilon_2-\epsilon_1)(N+1)/(N-1)$, 
with second order corrections due to the hopping term.
Asymmetric I-V characteristics with NDR peaks appearing at different voltages
in the forward and reverse bias directions is also a very notable point of 
this calculation.
We have also considered different spatial variations of the bias, having
larger gradients at the electrodes (as described in section II).
As seen from Figure \ref{IV_profile}, we observe that, whenever there is a 
strong spatial dependence of
the bias profile, the closing in of the HOMO and LUMO levels is more pronounced,
and the NDR sharper. A profile with almost no spatial variation can even
result in a no NDR situation.

\begin{figure}
\centering
\includegraphics[scale=0.5]{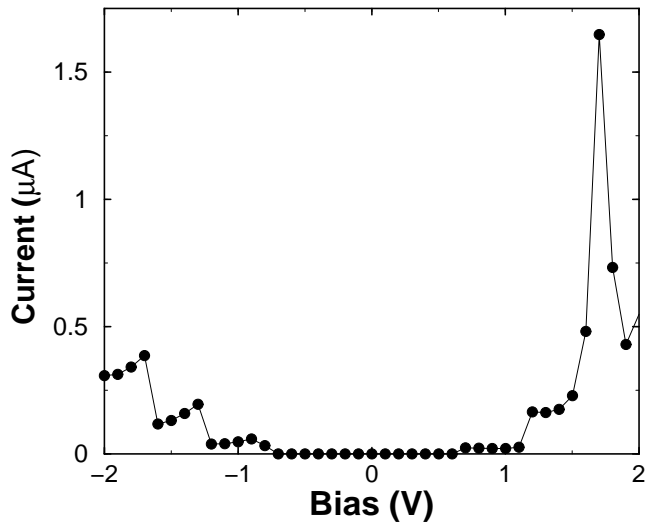}
\caption{ \label{BLA_DA} The I-V characteristics for the 20 site system with 
$t=1 eV$ and $\delta=0.2 eV$, where $\delta$ is the bond-alternation parameter.
The system also has a donor and acceptor at symmetric sites
4, 17 with energies $\epsilon_4=0.5 eV$ and  $\epsilon_{17}=-0.5 eV$.}
\end{figure}

In the long-chain limit, there is an implicit relation between variation
in on-site energies and the bond-length alternation (BLA). This is because
in the N $\rightarrow$ $\infty$ limit, the Fourier transformations of both
the diagonal (on-site) and off diagonal (hopping) terms yield
the same k (wavevector) components. The Tour 
molecules, as discussed before have both BLA (conjugation) together with 
donor and acceptor groups. We have already explained how
BLA which dimerizes the system, gives rise to NDR.
Interestingly, we find that acceptor ($+\epsilon$) and donor ($-\epsilon$) 
at some positions together with explicit dimerization also
causes NDR and asymmetric I-V, as shown in Figure \ref{BLA_DA}. 
Hence, we point out that, whether it is the 
explicit dimerization or two-sublattice structure, coupled with the voltage
profile, it induces interchange of symmetry together with Landau quasi-degeneracy of 
the low-lying levels, leading to NDR.

\section{ Finite size correlated insulators}
Having looked at the phenomenon of NDR in finite systems which are
inherently insulating, either due to Peierls distortion or because of
the presence of substituents, we naturally arrived at the question of
whether such insulator-metal-insulator transitions would survive in systems 
which are insulating due to the presence of strong electron-electron 
interactions.
Low-dimensional systems are almost always insulators, due to one or
all of the reasons mentioned above and are
commonly described by Peierls, Hubbard or related Hamiltonians 
\cite{ssh1,Hubbard}. It is well-known that the 
ground state of a half-filled Hubbard system is a Mott insulator,
and the effect of electric field on such a system has generated much 
interest in recent years due to the practical interest in tuning 
their dielectric and piezoelectric properties.
In our work detailed below \cite{Sudipta_JPCM}, we investigate the
effect of electric field on finite size correlated insulators,
using the exact diagonalization method 
and obtain the
ground and low-energy eigen states for various system sizes. 
We consider a one dimensional chain described by the Hubbard Hamiltonian,
\begin{eqnarray}
H &=& \sum\limits_{i}t(a^\dagger{_i}a_{i+1}+h.c) 
+ U \sum\limits_{i}n_{i\uparrow}n_{i\downarrow}
\end{eqnarray}

\noindent where t is the hopping term, U is the Hubbard term. We set 
$t=1$ as the unit of energy. The external electric field ($W$) applied on the 
system, has the form of a ramp potential, 
adds an extra term $\sum\limits_{i}W_{i}a^\dagger{_i}a{_i}$ 
to the above Hamiltonian. We perform calculation for various values of
$U$ and $W$.
\begin{figure}
\centering
\includegraphics[scale=0.4, angle=270] {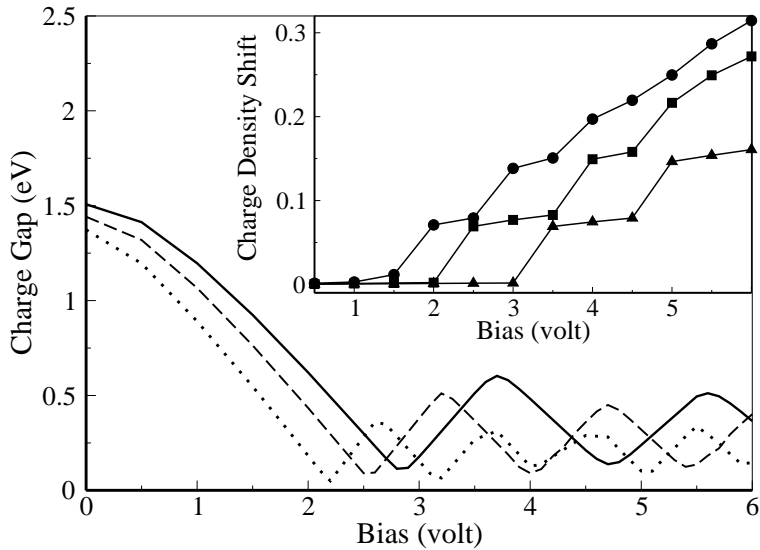}
\caption{\label{MI_fig1} Charge gap vs bias for $N=26$(solid line), $30$(dashed line)
and $40$(dotted line) for $U=4$. Inset shows the average charge density shift 
per site as a function of bias for a system, $N=30$ with $U=3(circle)$, 
$U=4(square)$ and $U=5(triangle)$.}
\end{figure}

The ground state of the Hamiltonian with nonzero $U$ is a spin-density 
wave insulator with one electron at every site \cite{Sumit}. To understand
the effect of the external field on this system, the charge excitation gap 
defined as the difference between the energy required to add ($\mu_+$) and 
remove ($\mu_-$) electrons from the ground state \cite{insulator}, i.e,
$\Delta_{charge}=\mu_{+}-\mu_{-}$ is calculated.
Here $\mu_{+}=E(N+1)-E(N)$ and $\mu_{-}=E(N)-E(N-1)$. $E(N)$, 
$E(N+1)$ and $E(N-1)$ are the energies of the half-filled system and
the systems with one extra and one less electron respectively.
We plot in Figure \ref{MI_fig1} this charge gap as a function of bias for 
different system sizes, for some representative value of $U$. It can 
be clearly seen that the charge gap shows an oscillation with bias, going 
through a number of minima and maxima. 

To understand the underlying reasons
for this oscillation, we calculate the average charge density shift per site 
as a function of bias for various system sizes with several $U$ values. In the inset 
of Figure \ref{MI_fig1}, we present the charge density shift per site for the half-filled 
state of a finite chain with $N=10$ for three representative values of
$U$. At zero bias, the ground state charge density at every site of the 
system is the same and it continues to remain so up to the bias
corresponding to the first minimum of the charge gap 
($\Delta_{charge}$). However, after that, it shows a large shift in 
the direction of bias, giving rise to charge inhomogeneities.
The external bias tends to shift the charge densities towards
one electrode with the nullification of $U$ at the first $\Delta_{charge}$ 
minimum. However, beyond this, an increase in bias results in further hopping 
of charges leading to double occupancy of more sites, with electron repulsion 
overwhelming the kinetic stabilization, thereby increasing the energy gap. 
Further increase of bias nullifies this effective repulsion, resulting in the 
next charge gap minimum. Hence, such a variation in charge gap resulting in 
near-metallic (charge gap is not zero) behavior in various bias regions is due 
to the interplay of the Hubbard repulsion, finite system size and the spatial 
gradient of the external bias. Interestingly, as can be seen from the figure, with 
increase in the size, the magnitude of bias corresponding to the first 
$\Delta_{charge}$ minimum reduces and the periodicity of occurrence of
successive minima thereafter also narrows down.
It is also evident from the inset that an increase in $U$
requires a higher bias to shift the charge density, 
and thus the bias corresponding to the first $\Delta_{charge}$ 
minimum also increases.


For a simple understanding on the above breakdown phenomenon, we consider
a 2 site Mott-insulator in presence of bias. With ramp potential, the
first and the second site experiences the bias, $-\frac{W}{6}$ and $+\frac{W}{6}$
respectively. Out of $4^2$ states, (a) the lowest energy of the one-electron
states is $E_1 = -(\sqrt{(36t^{2}+W^{2}))}/6$, while (b) for the 3 electron
states it is, $E_3 = (6U - \sqrt{(36t^{2}+W^{2}))}/6$. These bias dependent
energies are stabilized with increase in bias, however, as expected, their 
energy difference is $U$ for any $W$. The ground state of the half-filled 
system comprising of four basis states: (i) $|\uparrow\downarrow$ $0$ $>$,
(ii) $|\uparrow$ $\downarrow$ $>$, (iii) $|\downarrow$ $\uparrow$ $>$ and
(iv) $|0$ $\uparrow\downarrow$ $>$. For $W=0$, the ground state consists of
singly occupied states (ii \& iii), with second order ($t^2/U$) contribution
from double occupancy site states (i \& iv). However, as the bias increases,
the state with double occupancy (state i) starts to contribute to the ground 
state energy (of the half-filled system), $E_2$. It is because the external bias 
nullifies the Hubbard repulsion, leading to mixing of double and single occupancy states. This 
results in an increased hopping of electrons and hence a closing of the 
repulsion induced charge gap at $W_c$. Beyond $W_c$, the state with double 
occupancy (state i) starts gaining prominence and the energy $E_2$ stabilizes 
with bias. This results in an increasing value of $\mu_{+}$ and a decreasing 
value of $\mu_{-}$ and hence a rise in the charge gap leading to an insulating 
state again. Although in this simple 2-site model, the states with one-electron 
never faces the effect of $U$, it could, however, capture the essence of the 
insulator-near metal transition, albeit qualitatively.

These insulator-near metal-insulator transitions strongly indicate that
the current through these systems would show a number of NDR
peaks\cite{anirban}.

\section{NDR in double quantum dots in the \\ coulomb blockade regime}
In this section, we focus on the NDR phenomenon that has 
recently been observed in double quantum dots in the 
low temperature, weak molecule-electrode coupling limit, also
known as the coulomb blockade regime \cite{Ono}. 
Theoretical studies of NDR in this
regime have also started gaining a lot of prominence in
recent times \cite{Hettler1,Hung,Thielmann}. 
Mean field descriptions are known to usually fail here,
as electron charging  energies are very high as compared to the broadenings due 
to the electrode coupling \cite{Datta_MB,Bhaskaran}.
Also, since these methods, combined with standard non-equilibrium
Greens function (NEGF) treatment of transport are perturbative in the interaction
parameter, they cannot capture the transitions between the spectrum of 
neutral and excited
states, which can lead to a variety of interesting 
features in the I-V characteristics.
The formalism that has now come to be used widely to capture molecular transport
in the coulomb blockade regime is the master or rate equation method 
\cite{Hettler,Beenakker}. Here, we use this formalism to study a system consisting 
of a donor and acceptor in the coulomb blockade regime.
Taking cue from our mean-field transport studies on such systems
which showed NDR behavior, as discussed before \cite{lakshmi_PRB}, here we explore 
the role of strong correlations in determining their transport in the single electron 
charging regime \cite{lakshmi_KE_PRB}. 
The rate equation formalism describes transport through a correlated system
with many-body eigenstates.
The presence of coulomb interactions results in occupation probabilities
of each many body state that cannot be factorized as the product of the occupation
probabilities of each single electron level. Hence, in this case, the full
rate-equation
problem, where the occupation probability of each many-body state is treated as an
independent variable is solved, neglecting off-diagonal coherences. 
In this method, the transition rate $\Sigma_{{s^\prime}\rightarrow{s}}$
from the many-body state $s^\prime$ to $s$, involving transitions between states
differing by one electron, is calculated up to linear order 
in $\Gamma$ (which is the bare
electron tunneling rate between the system and the electrode), using Fermi's golden
rule (from second-order perturbation theory) as \cite{Hettler,Beenakker},

\begin{figure}
\centering
\includegraphics[width=0.7\columnwidth]{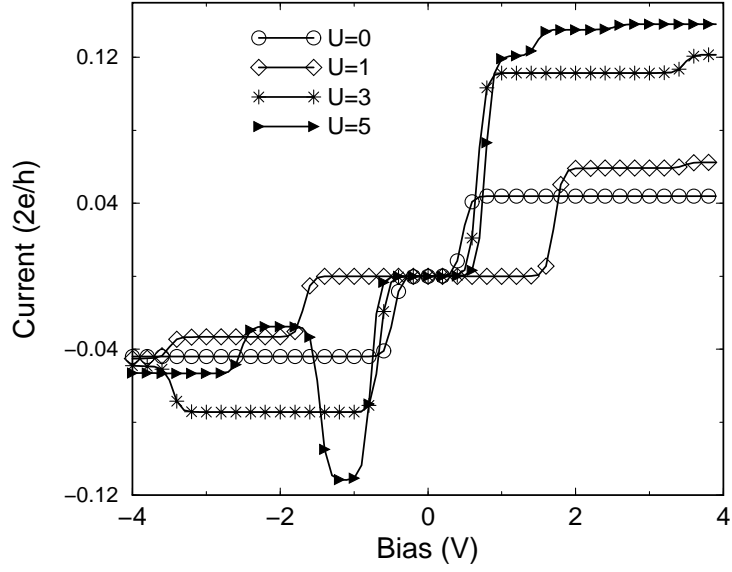}
\caption{ \label{2site_NDR} The I-V characteristics of the 2 site donor-acceptor
system obtained using the rate equation approach, for various values of
the Hubbard parameter $U$. Here, $\epsilon_1=-\epsilon_2= 2.0$eV, $t=1.0$eV.}
\end{figure}

\bear
\Sigma_{{s^\prime}\rightarrow{s}}^{L+}&=&\Gamma f_L(E_s-E_s^\prime) \sum_{\sigma} |<s|C^\dag_{1\sigma}|s^\prime>|^2 \\ 
\Sigma_{{s^\prime}\rightarrow{s}}^{R+}&=&\Gamma f_R(E_s-E_s^\prime) \sum_{\sigma} |<s|C^\dag_{N\sigma}|s^\prime>|^2 \nonumber
\eear
\noindent with a corresponding equation for 
$\Sigma_{{s} \rightarrow {s^\prime}}^{L-}$ and
$\Sigma_{{s} \rightarrow {s^\prime}}^{R-}$ obtained 
by replacing $f_{L,R}(E_s-E_s^\prime)$ by
$(1-f_{L,R}(E_s-E_s^\prime))$. Here, $+/-$ correspond to the 
creation/annihilation of an electron
inside the molecule due to electron movement from/to left (L) or right (R) electrode.
We have also assumed that the creation and annihilation happen only at 
the terminal sites.
The total transition rate is then obtained as,
$\Sigma_{{s} \rightarrow {s^\prime}}=\Sigma_{{s} \rightarrow {s^\prime}}^{L+}+\Sigma_{{s} \rightarrow {s^\prime}}^{R+}+\Sigma_{{s} \rightarrow {s^\prime}}^{L-}+\Sigma_{{s} \rightarrow {s^\prime}}^{R-}$.
The non-equilibrium probability $P_s$ of occurrence of each many-body state $s$, is obtained
by solving the set of independent rate equations defined by
$\dot{P_s}=\sum_{s^\prime}\Sigma_{{s^\prime} \rightarrow s}P_{s^\prime}-\Sigma_{{s} \rightarrow {s^\prime}} P_s$
\noindent through the stationarity condition $\dot{P_s}=0$ at steady state. This
results in a
homogeneous set of equations of the size of the Fock space. Taking advantage
of the normalization condition $\sum_s{P_s}=1$, we obtain linear equations,
which can be solved using well-known linear algebraic methods.
The steady state probabilities are then used to obtain the terminal current
as ($\alpha=L/R$ electrode),
\bear
I_{\alpha}={e \over \hbar} \sum_{s,s^\prime}\Sigma_{{s^\prime} \rightarrow s}^{\alpha+}P_{s^\prime}-\Sigma_{{s} \rightarrow {s^\prime}}^{\alpha-} P_s
\label{cureq}
\eear

There have been previous
theoretical studies on donor-acceptor double dot systems where
strong rectification has been observed \cite{Cuniberti}, and others which showed
NDR with variation in the molecule-electrode coupling \cite{Thielmann, Aghassi} or
due to a detuning of the molecular levels \cite{Aghassi}. Another recent
study has attempted to establish conditions obeyed by the parameters
involved, to observe a collapse in the current \cite{Bhaskaran_generic}. 
However, in our study, we find that some subtle changes in the parameters do result 
in NDR peaks in regimes where the conditions may not be followed strictly.
Our Hamiltonian for the two site system is written as,
\bear
H =  \sum_{i=1}^2\ \epsilon_i(a^{\dag}_{i}a_{i})+ \sum_{\sigma} -
t(a^{\dag}_{1\sigma}a_{2\sigma}+hc)
 +  U \sum_{i=1}^2 n_{i\uparrow}n_{i\downarrow} \nonumber
\eear

\noindent where $t$ is the hopping strength between the donor and acceptor, 
$\epsilon_{1,2}$
are the on-site energies and U is the Hubbard interaction between 
electrons at the same site. 
For obtaining current, for every value of $U$, the Fermi energy ($E_F$) is 
chosen as the gate bias value which
ensures that the two electron state is the ground state.
The Fermi energy is also placed in such a way that we observe
transitions from the ground state to the state with one less electron.
Results however, would look similar for transitions
to the state with one more electron.

\begin{figure}
\centering
\includegraphics[width=0.6\columnwidth]{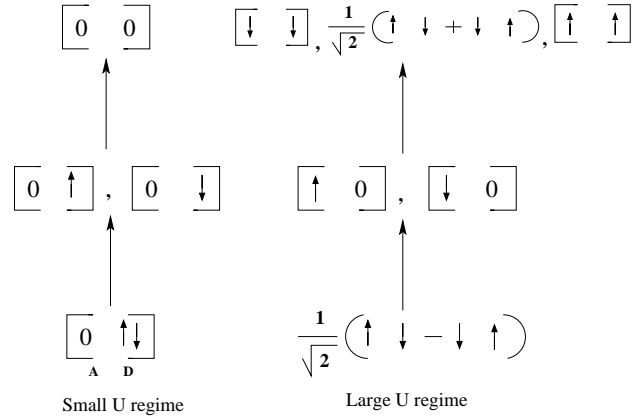}
\caption{\label{schematic_KE} A schematic describing the transitions between the 
states of the donor-acceptor system, in the small and large U regimes.}
\end{figure}
In Figure \ref{2site_NDR}, we plot the current-voltage characteristics of
the system, for various values of
the Hubbard parameter. As can be seen clearly, low values of $U$
result in step-like features in the I-V, due to transitions from the
2 electron (2$e^-$) singlet ground state to the 1 electron (1$e^-$) doublet states and then
to the state with no electron. Interestingly, with increase in $U$, a rise
and fall in current (a NDR feature) is observed for negative values of bias.
An analysis of the probabilities shows that this
happens when the second jump in the system occurs from the 1$e^-$ 
doublet to
a higher excitation of the 2$e^-$ state, namely the triplet
states, instead of to the state with no electron. This is because, when U is small,
the ground state gives higher preference to the state with 2$e^-$ of 
opposite spins
at the site with lower on-site energy. This allows annihilation of an electron by the
electrode followed by one more annihilation leading to a transition
from the 2$e^-$ singlet to the 1$e^-$ doublet to the
state with 0 electron. When U increases, however, the ground state
gives more weightage to the state with 1$e^-$ at the donor
and one at the acceptor (see Figure \ref{schematic_KE}). 
This allows for one annihilation from the ground state to the
1$e^-$ doublet state, followed by a creation from the same electrode
to the 2$e^-$ triplet state, which has the same energy as the 0 electron state.
Since the current at any electrode is calculated at steady state as the
difference between the outgoing and incoming current (see Eqn.\ref{cureq}),
this transition results in a reduction in current leading to the
negative differential resistance peak (also known as spin-blockade). 

The I-V is asymmetric because of the inherent
asymmetry in the system comprising of a donor and an acceptor. This becomes
very apparent for larger U values for which NDR appears only at one
polarity of the applied bias. This is because, the transition from
1$e^-$ doublet to the 2$e^-$ triplet through a creation at
the acceptor is less feasible than
at the donor.

\section{Conclusions}

In conclusion, we have attempted to present the effects of electric filed
on molecular systems between macroscopic electrodes, to gain an 
understanding of some of the factors that determine how the I-V 
characteristics of a nanoscale system would look. As a first
step, we have elaborated the importance of the potential profile i.e the
way the external applied field would fall across the molecule,
and how the presence of electron correlations in the molecule
can drastically change its shape, thereby changing the device
response along-with it. Furthermore, highlighting the negative differential
resistance (NDR) phenomena in molecules that has captivated the attention
of researchers due to its immediate application in switching devices,
we have shown that the presence of a two sublattice structure,
caused either by a lattice distortion or substituents with a large
dipolar strength, can result in NDR, when the external field has a strong 
spatial dependence. We find that the ratio 
$t$ (hopping integral) : $\epsilon_2-\epsilon_1$ (measure of the dipolar character), 
is very crucial in determining the nature of the I-V characteristics.
Three features, namely (1) the critical bias (2) the sharpness of the 
NDR peak and (3) the extent of asymmetry in the I-V curves are sensitive to this ratio. 
Although a common thread between all the known factors 
governing NDR is yet to be understood, we think that our reasoning even encompasses 
other existing explanations such as bias driven conformational changes, reduction of 
the acceptor to donor, as well as the bias driven changes in electrode-molecule coupling, 
all of which would equivalently result in a change in the electronic structure caused
by the external bias. Our study also highlights the importance of a strong spatial
dependence of the potential profile to obtain a sharp NDR peak. A natural question of
whether such an NDR will survive when the system is insulating due to presence of
electron-electron interactions in them, led us to study correlated insulators in the
presence of field, where too we have shown that insulator-near metal-insulator transitions
can be brought about by the spatially varying external bias, thus opening up possibilities
for switching in these systems. Additionally, we have elucidated how a variation of these 
interaction strengths in double quantum dots can result in device responses varying from
normal coulomb blockade to NDR.

It is mandatory to say at this point that most of our studies have assumed tunneling
as the mechanism of transport and calculations have been performed in the coherent 
transport regime. However, the field itself is rich in studies on incoherent as well
as activated transport characteristics. Experimental observation of phase loss in DNA junctions 
\cite{DNA_inco} and theoretical studies based on the Buttiker probe \cite{Buttiker} 
or generalized master equations \cite{GME} are just some of the efforts in this 
direction, and a proper understanding of decoherence in condensed phase transport is 
still a challange \cite{Ratner_rev}. Electron-electron interections, which in 
molecular transport are the source of Coulomb blockade phenomena, have been explored in some
of the later parts of this review, however, there have been many observations of
both the Coulomb blockade and the Kondo effect in molecular junctions,
and very often they are accompanied by vibrational features as well, which
make transport phenomena more interesting with advanced technological applications \cite{CB_Kondo}. 
Theoretical approaches based on the equation of motion method \cite{EOM} or Fock space formalism 
\cite{Pati_IEEE} have also begun 
to be appear to gain understanding of this regime. Although, we have discussed mainly
about one of the technologically useful non-linear phenomena, namely, Negative differential resistance, 
other phenomena such as multistability and hysteresis, 
which arise out of structural changes in a molecule during transport together with inelastic processes 
due to interactions at different energy and length scales in organic, inorganic or hybrid systems still remain the 
subjects of active current research \cite{hyst1,amlan,hyst2,hyst3}.

\section{Acknowledgements}

SD acknowledges the CSIR for the research fellowship and SKP acknowledges the research
support from DST and CSIR, Govt. of India.

\end{document}